\documentclass[dvips]{article}
\usepackage{supertabular,lscape,epsfig}
\usepackage{amssymb}
\usepackage{amsmath}

\usepackage{pslatex}

\DeclareSymbolFont{ppa}{OT1}{ppl}{m}{it}
\DeclareMathSymbol{\vv}{\mathalpha}{ppa}{'166}

\begin{document}

\newcommand{\dd}{\,{\rm d}}
\newcommand{\ie}{{\it i.e.},\,}
\newcommand{\etal}{{\it et al.\ }}
\newcommand{\eg}{{\it e.g.},\,}
\newcommand{\cf}{{\it cf.\ }}
\newcommand{\vs}{{\it vs.\ }}
\newcommand{\zdot}{\makebox[0pt][l]{.}}
\newcommand{\up}[1]{\ifmmode^{\rm #1}\else$^{\rm #1}$\fi}
\newcommand{\dn}[1]{\ifmmode_{\rm #1}\else$_{\rm #1}$\fi}
\newcommand{\upd}{\up{d}}
\newcommand{\uph}{\up{h}}
\newcommand{\upm}{\up{m}}  
\newcommand{\ups}{\up{s}}
\newcommand{\arcd}{\ifmmode^{\circ}\else$^{\circ}$\fi}
\newcommand{\arcm}{\ifmmode{'}\else$'$\fi}
\newcommand{\arcs}{\ifmmode{''}\else$''$\fi}
\newcommand{\MS}{{\rm M}\ifmmode_{\odot}\else$_{\odot}$\fi}
\newcommand{\RS}{{\rm R}\ifmmode_{\odot}\else$_{\odot}$\fi}
\newcommand{\LS}{{\rm L}\ifmmode_{\odot}\else$_{\odot}$\fi}

\newcommand{\Abstract}[2]{{\footnotesize\begin{center}ABSTRACT\end{center}
\vspace{1mm}\par#1\par   
\noindent
{~}{\it #2}}}

\newcommand{\TabCap}[2]{\begin{center}\parbox[t]{#1}{\begin{center}
  \small {\spaceskip 2pt plus 1pt minus 1pt T a b l e}
  \refstepcounter{table}\thetable \\[2mm]
  \footnotesize #2 \end{center}}\end{center}}

\newcommand{\TableSep}[2]{\begin{table}[p]\vspace{#1}
\TabCap{#2}\end{table}}

\newcommand{\FigCap}[1]{\footnotesize\par\noindent Fig.\  %
  \refstepcounter{figure}\thefigure. #1\par}

\newcommand{\TableFont}{\footnotesize}
\newcommand{\TableFontIt}{\ttit}
\newcommand{\SetTableFont}[1]{\renewcommand{\TableFont}{#1}}

\newcommand{\MakeTable}[4]{\begin{table}[htb]\TabCap{#2}{#3}
  \begin{center} \TableFont \begin{tabular}{#1} #4
  \end{tabular}\end{center}\end{table}}

\newcommand{\MakeTableSep}[4]{\begin{table}[p]\TabCap{#2}{#3}
  \begin{center} \TableFont \begin{tabular}{#1} #4
  \end{tabular}\end{center}\end{table}}

\newenvironment{references}%
{
\footnotesize \frenchspacing
\renewcommand{\thesection}{}
\renewcommand{\in}{{\rm in }}
\renewcommand{\AA}{Astron.\ Astrophys.}
\newcommand{\AAS}{Astron.~Astrophys.~Suppl.~Ser.}
\newcommand{\ApJ}{Astrophys.\ J.}
\newcommand{\ApJS}{Astrophys.\ J.~Suppl.~Ser.}
\newcommand{\ApJL}{Astrophys.\ J.~Letters}
\newcommand{\AJ}{Astron.\ J.}
\newcommand{\IBVS}{IBVS}
\newcommand{\PASP}{P.A.S.P.}
\newcommand{\Acta}{Acta Astron.}
\newcommand{\MNRAS}{MNRAS}
\renewcommand{\and}{{\rm and }}
\section{{\rm REFERENCES}}
\sloppy \hyphenpenalty10000
\begin{list}{}{\leftmargin1cm\listparindent-1cm
\itemindent\listparindent\parsep0pt\itemsep0pt}}%
{\end{list}\vspace{2mm}}
 
\def\TYLDA{~}
\newlength{\DW}
\settowidth{\DW}{0}
\newcommand{\dw}{\hspace{\DW}}

\newcommand{\refitem}[5]{\item[]{#1} #2%
\def\REFARG{#3}\ifx\REFARG\TYLDA\else, {\it#3}\fi
\def\REFARG{#4}\ifx\REFARG\TYLDA\else, {\bf#4}\fi
\def\REFARG{#5}\ifx\REFARG\TYLDA\else, {#5}\fi.}

\newcommand{\Section}[1]{\section{#1}}
\newcommand{\Subsection}[1]{\subsection{#1}}
\newcommand{\Acknow}[1]{\par\vspace{5mm}{\bf Acknowledgements.} #1}
\pagestyle{myheadings}

\newfont{\bb}{ptmbi8t at 12pt}
\newcommand{\xrule}{\rule{0pt}{2.5ex}}  
\newcommand{\xxrule}{\rule[-1.8ex]{0pt}{4.5ex}}  
\def\thefootnote{\fnsymbol{footnote}}
\newcommand{\uprule}{\rule{0pt}{2.5ex}}
\newcommand{\douprule}{\rule[-2ex]{0pt}{4.5ex}}
\newcommand{\dorule}{\rule[-2ex]{0pt}{2ex}}

\begin{center}
{\Large\bf
Evolution of Cool Close Binaries - Approach to Contact}
\vskip1.7cm
{\bf 
K.~~ S~t~\c e~p~i~e~\'n}
\vskip7mm
{Warsaw University Observatory, Al.~Ujazdowskie~4, 00-478~Warszawa, Poland\\
e-mail:kst@astrouw.edu.pl}
\end{center}

\Abstract{As a part of a larger project, a set of 27 evolutionary models of
cool close binaries was computed under the assumption that their evolution
is influenced by the magnetized winds blowing from both components. Short
period binaries with the initial periods of 1.5, 2.0 and 2.5 d were
considered. For each period three values of 1.3, 1.1 and 0.9 $M_{\odot}$
were taken as the initial masses of the more massive components. The
initial masses of the less massive components were adjusted to avoid
extreme mass ratios.

Here the results of the computations of the first evolutionary phase are
presented, which starts from the initial conditions and ends when the more
massive component reaches its critical Roche lobe. In all considered cases
this phase lasts for several Gyr. For binaries with the higher total mass
and/or longer initial periods this time is equal to, or longer than the
main sequence life time of the more massive component. For the remaining
binaries it amounts to a substantial fraction of this life time.

From the statistical analysis of models, the predicted period distribution
of detached binaries with periods shorter than 2 d was obtained and
compared to the observed distribution from the ASAS data. An excellent
agreement was obtained under the assumption that the period distribution in
this range is determined solely by magnetic braking (MB) \ie the mass and
angular momentum loss due to the magnetized winds, as considered in the
present paper. This result indicates, in particular, that virtually all
cool detached binaries with periods of a few tenths of a day, believed to
be the immediate progenitors of W UMa-type stars, were formed from young
detached systems with periods around 2-3 d. MB is the dominant formation 
mechanism of cool contact binaries. It operates on the time scale of 
several Gyr rendering them rather old, with age of 6-10 Gyr.

The results of the present analysis will be used as input data to
investigate the subsequent evolution of the binaries, through the mass
exchange phase and contact or semi-detached configuration till the ultimate
merging of the components.}

{Stars: activity -- binaries: close -- Stars:
evolution -- Stars: late-type -- Stars: rotation}

\Section{Introduction}

W~UMa-type binaries are contact binaries consisting
of two cool stars surrounded by a common envelope lying between the inner
and outer Lagrangian zero velocity equipotential surfaces, called also the
Roche lobes (Mochnacki 1981). In spite of different component masses they
possess nearly identical mean surface brightnesses. The more
massive, primary component is a main sequence (MS) star whereas the
secondary is oversized compared to its expected MS radius.

Kuiper (1941) noted that contact binaries with unequal zero-age components
cannot exist in equilibrium because radii of the components must fulfill
two, mutually contradictory conditions: one resulting from the mass-radius
relation for zero-age stars and the other relating sizes of the Roche
lobes, identical in this case with stellar sizes, to stellar masses. The
fact that contact binaries are nevertheless observed is known as ``Kuiper
paradox''. Lucy (1968) noted that the Kuiper paradox can be solved by
assuming that both zero-age components (each in the thermal equilibrium) 
have a common convective envelope
with the same adiabatic constant. However, detailed calculations showed
that realistic models can exist only within a narrow range of component
masses, contrary to what was observed. Later Lucy (1976) considered a
binary configuration which is not in equilibrium. A similar model was
developed by Flannery (1976). The model is known as Thermal Relaxation
Oscillation (TRO) model. According to it each component of the binary is
out of thermal equilibrium and its size oscillates around the inner Roche
lobe but the whole binary is in the global thermal equilibrium. The energy
is transported from primary to secondary via a turbulent convection which
results in equal entropies of both convective envelopes, hence equal
surface brightnesses.

TRO model explains in a very elegant way two basic observational facts
about W~UMa-type stars: the Kuiper paradox, i.e. the geometry of the binary
in which primary is an ordinary MS star and secondary is also a MS star but
swollen to the size of its Roche lobe by the energy transfer, and
equal apparent effective temperatures of both components resulting in a
characteristic light curve with two equal minima. It encounters,
however, several difficulties. In particular, some of its basic assumptions
seem to be incorrect and some of its predictions are at odds with the
observations. 

TRO model predicts that a binary oscillates between two states:
contact -- when both stars fill their respective Roche lobes and there is a
net mass flow from the secondary to the primary, and semidetached -- when
the primary still fills its Roche lobe but the secondary is within its
Roche lobe and the mass flows from the primary to the secondary (Lucy
1976). Robertson and Eggleton (1977) argued that the time scales of
both states should be close to the thermal time scales of both
components. For most W UMa-type stars the mass ratio is rather moderate, 
about 0.5, so the time scales should be comparable to
one another. This is at odds with observations showing that contact (or
near-contact) binaries within the same parameter range as genuine contact
binaries but with distinctly different depths of minima are quite rare
(Rucinski 1998). Additional effects, like stellar evolution and/or angular
momentum loss (AML) can influence the ratio of both time scales reducing
the duration of the semidetached state (Robertson and Eggleton 1977,
Rahunen 1983, Yakut and Eggleton 2005, Li, Han and Zhang 2005). In fact, a
binary can remain in the contact state all the time if AML rate is high
enough but then its lifetime as a contact binary must be as short as $\sim
10^8$ years.  Such binaries would be rare in space, contrary to observations
(Webbink 2003).

A recent photometric sky survey ASAS (Pojmanski 2002) detected several
thousand eclipsing binaries with periods shorter than 1 day (Paczy\'nski
\etal 2006). Their analysis showed a significant proportion of
semi-detached systems. This would seemingly solve the problem and support
TRO model. However, a closer look at the sample shows that among stars with
periods shorter than 0.45 day there is very few semi-detached binaries
(Pilecki, 2010) whereas a great majority of contact stars has periods
within this range (Rucinski 2007). On the other hand, semi-detached
binaries are commonly observed among binaries with periods between 0.7 and
1 day where fewer contact systems occur. In addition, one should keep
in mind that, owing to the larger expected light variation amplitude of
semi-detached binaries, they are preferentially detected over small mass
ratio contact binaries with shallow minima (Rucinski 2006). So, the
problem still remains.

Section 2 contains a short description of the essentials of a new model of
the origin of cool contact binaries, suggested by the present author.  In
Section 3 the assumptions, equations and other details of the model are
given and Section 4 presents the initial parameters of the considered
binaries together with the results of the model calculations. In Section 5
a comparison of the model data with observations is given and the most
important uncertainties of the model are discussed. Section 6 contains the
main conclusions of the paper and plans for continuation of this
investigation.

\section{Essentials of a New Model}

The original TRO model has been developed for zero-age stars.
However, the numerical simulations of the binary formation favored an early
fragmentation of a protostellar cloud (Boss 1993, Bonnel 2001, Machida
\etal 2008)) rather than fission. The observations support this
conclusion. We do not observe binaries with orbital periods shorter than
one day in young stellar clusters or among T Tauri-type stars but they are
very numerous in old open clusters with an age exceeding 4-5 Gyr and in
globular clusters (Mathieu 1994, Kaluzny and Rucinski 1993, Rucinski 1998,
2000). Large numbers of W~UMa-type stars have been observed in the galactic
bulge (Szyma\'nski, Kubiak and Udalski 2001). The analysis of kinematics of
these stars in the solar vicinity showed that they are several Gyr old
(Guinan and Bradstreet 1988, Bilir \etal 2005). The time interval of 5 to
13 Gyr is long enough for a primary with mass between 1.3 and 0.9
$M_{\odot}$ to complete its MS evolution. We should thus ubiquitously
observe W~UMa-type stars with components possessing hydrogen-depleted
cores. Such a star expands rapidly when moving towards the red giant branch
and its evolution must very strongly influence the fate of the whole
binary.

Based on these facts a new model of a cool contact binary has been
developed by the present author (St\c epie\'n 2004, 2006a, 2006b, 2009).
The model assumes that contact binaries originate from young cool close
binaries with initial orbital periods close to a couple of days. Both
components possess subphotospheric convective zones necessary for
generating surface magnetic fields. Such stars show magnetic activity 
with intensity increasing with the
increase of rotational angular velocity. Assuming the synchronization of
rotation with orbital period we expect the activity level of each component
to be at the very high, so called ``saturation'' level. The magnetic
activity drives stellar winds carrying away the spin angular momentum
(AM). Because of synchronization, this loss results in a decrease of the
orbital AM. The orbit tightens and the components approach each other until
the more massive component fills the critical Roche
surface. 

Because of the ambiguity of the term ``primary'' which is
sometimes used to describe the more massive component and sometimes the
component with the higher surface brightness we will identify the
components as star ``1'' and star ``2'' according to their initial mass:
$M_1$ denotes initially the more massive component and $M_2$ initially the
less massive component, so that initial mass ratio $q_{\mathrm{init}}
\equiv M_1/M_2 \ge 1$ whereas after mass ratio reversal $q \le 1$.  For
$M_{\mathrm{1,init}} \approx 1.3 M_{\odot}$ and the initial period of 2 d
the time scale for AML is close to the lifetime of the star ``1'' on MS so
the Roche lobe overflows (RLOF) begins when the star is close to, or beyond
the terminal age main sequence (TAMS) on the Hertzsprung-Russell (HR)
diagram and its radius increases quickly on the way to red giant
region. RLOF by star ``1'' results in the mass transfers to the component
``2''. The model assumes that the mass transfer proceeds in the same way as
in Algols, \ie after a possible phase of a rapid mass exchange resulting
in a loss of thermal equilibrium of both components and a transient
formation of the common envelope, both stars reach the thermal equilibrium
within a configuration with the mass ratio reversed. In binaries with the
mass ratio not far from unity star ``2'' approaches zero age main sequence
(ZAMS) because it has been fed with hydrogen rich matter and in binaries
with the low mass ratio it simply stays close to ZAMS. Star ``1'' is
evolutionary over sized because its core is hydrogen depleted. After the
phase of the rapid mass exchange is over, star ``1'' continues expansion
due to increase of its helium core which results in a further mass transfer
although at a much reduced rate. The mass transfer rate in this
evolutionary phase depends on two effects acting in opposite directions:
mass transfer results in the orbit expansion whereas AML tightens
it. Interplay between these two processes determines the rate of evolution
of a contact binary towards the extreme mass-ratio configuration which
should ultimately end up in merging of both components. In some binaries
star ``2'' may become massive enough that its evolutionary expansion may
become important. In particular, if it reaches TAMS, the common envelope
will overflow the outer critical Roche surface and a violent
coalescence of both components is expected.

Models published hitherto by the present author were calculated for a few
selected component masses and the initial orbital period of 2 days. Now the
results of calculations of 27 models with different component masses,
covering the broad range of mass ratios and with three different initial
periods: 1.5, 2.0 and 2.5 d will be discussed. Analysis of their evolution
provides data which can be compared with observations of close binaries --
both with individual systems and with statistical data. The present paper
discusses in detail the first evolutionary phase of the modeled binaries:
the approach to contact.

\Section{Assumptions, equations and details of the model}

According to our assumptions cool contact binaries are formed from close
detached binaries which lose AM via the magnetized wind. AML loss is
sufficiently efficient to form a contact binary within several Gyr if both
components possess magnetized winds. Therefore we consider only binaries
with initial component masses less than the limit for the occurrence of the
subphotospheric convective zone. Let us fix this mass at 1.3 $M_{\odot}$ so
that the total initial mass of a binary is less than 2.6 $M_{\odot}$. We
apply the Roche model for the description of the orbital parameters of the
binary. The first model equation is the third Kepler law

\begin{equation}
P = 0.1159a^{3/2}M^{-1/2}\,,
\end{equation}

where $M = M_{\rm{1}} + M_{\rm{2}}$ is the total mass, $a$ -- semi axis and
$P$ -- orbital period. Here masses and $a$ are expressed in solar units,
and period in days. We also need an expression for the total
angular momentum

\begin{equation}
H_{\rm{tot}} = H_{\rm{spin}} + H_{\rm{orb}}\,,
\end{equation}

where 

\begin{equation}
H_{\rm{spin}} = 7.13\times10^{50}(k^2_1M_1R^2_1 +k^2_2M_2R^2_2)P^{-1}\,,
\end{equation}

and

\begin{equation}
H_{\rm{orb}} = 1.24\times 10^{52}M^{5/3}P^{1/3}q(1+q)^{-2}\,,
\end{equation}

where $k_1^2$ and $k^2_2$ are gyration radii of both components and
 $H_{\rm{spin}}$ and $H_{\rm{orb}}$ are in cgs units. The approximations
 for critical Roche lobe sizes, $r_1$ and $r_2$, given by Eggleton (1983)
 are adopted

\begin{equation}
\frac{r_{\rm{1}}}{a} = \frac{0.49q^{2/3}}{0.6q^{2/3}+\ln{(1 + q^{1/3})}}\,,
\end{equation}

\begin{equation}
\frac{r_{\rm{2}}}{a} = \frac{0.49q^{-2/3}}{0.6q^{-2/3}+\ln{(1 +
q^{-1/3})}}\,,
\end{equation}
 
where $q = M_{\rm{1}}/M_{\rm{2}}$ is the mass ratio.  

For the considered binaries spin AM is nearly two
 orders of magnitude lower than orbital AM and can be neglected.

The empirical AML
rate of single, solar-type stars was found by St\c epie\'n (1995), based
on the observations of the rotation of stars slightly hotter than the Sun
(\ie stars with $<\,B-V\,> = 0.60$) in stellar clusters of different
age. Assuming that each component of a binary loses AM as a single star,
the AML rate of a contact binary was found. The resulting formula was
parameter-free, \ie the rate depends solely on the binary parameters. Later,
observational data on spin down of stars of different mass became available
which permitted to generalize the expression for AML rate of a close
synchronized binary with arbitrary mass and radius (St\c epie\'n 2006b,
Gazeas and St\c epie\'n 2008). The formula reads

\begin{equation}
\frac{{\rm d}H_{\rm{orb}}}{{\rm d}t} = -4.9\times 10^{41}
(R_1^2M_1 +R_2^2M_2)/P\,.
\end{equation}

Note that the above formula does not contain the exponential term appearing
in the expression for AML of a single star because its
value is close to 1 for rapid rotation (St\c epie\'n 2006b). In such a
limit the AML rate depends solely on the orbital period, stellar masses and
radii, as Eq.~(7) shows. The uncertainty of the numerical coefficient is of
the order of 50\%. For a binary with two solar components the formula
reduces to the same expression as derived in St\c epie\'n (1995). 

The observations of X-ray flux of single rapid rotators show a {\em
decrease} of the flux for rotation periods shorter than 0.4 d (Randich et
al. 1996, St\c epie\'n \etal 2001). This phenomenon is called {\em
supersaturation}. Assuming that supersaturation affects also the
magnetized wind, it was conservatively adopted that AML levels off for
rotation periods shorter than 0.4 days.

The magnetized wind carries away not only AM but also mass. Wood et
al. (2002) and Wood et al. (2005) determined mass loss rates of several
single, active stars of the solar type. A fit to their data gives in the
saturation limit (St\c epie\'n 2011) 

\begin{equation}
\dot M_{1,2} = -10^{-11}R_{1,2}^2\,,
\end{equation}

where mass loss rates are in  $M_{\odot}$/year and radii in solar
units. The relation applies for stars with $M_{1,2} \leq 1 M_{\odot}$. For
more massive stars $R_{1,2} \equiv 1$ is adopted.
Note that the above formula is also parameter-free. The
numerical coefficient is uncertain within the factor of two.

\begin{table}
\caption[]{Initial parameters of the computed models}
\begin{tabular}{ccccccccrcccc} \hline\hline

$M_1$ && $M_2$ && $q$ && $a$ && $H_{\rm{orb}}\quad$  && $R_1/r_1$ && $R_2/r_2$ \\
($M_{\odot}$) &&  ($M_{\odot}$) && ($M_1/M_2$) && ($R_{\odot}$) && ($\times
10^{51}$) && (percent) && (percent) \\
\hline
\hline
&&&&&&$P_{\rm init}$ = 2.5 d &&&&&& \\
\hline
\hline
1.3 && 1.1 && 1.18 && 10.37 && 17.97 && 30 && 28 \\
1.3 && 0.9 && 1.44 && 10.08 && 15.14 && 29 && 25   \\
1.3 && 0.7 && 1.86 &&  9.76 && 12.15 && 28 && 21 \\
1.1 && 0.9 && 1.22 &&  9.76 && 13.22 && 26 && 23 \\
1.1 && 0.7 && 1.57 &&  9.43 && 10.65 && 25 && 20 \\
1.1 && 0.5 && 2.20 &&  9.06 &&  7.91 && 25 && 16 \\
0.9 && 0.7 && 1.29 &&  9.06 &&  9.06 && 21 && 19 \\
0.9 && 0.5 && 1.80 &&  8.67 &&  6.77 && 21 && 16 \\
0.9 && 0.3 && 3.00 &&  8.23 &&  4.28 && 20 && 11 \\
\hline
\hline
&&&&&& $P_{\rm init}$ = 2.0 d &&&&&& \\
\hline
\hline
1.3 && 1.1 && 1.18 &&  8.94 && 16.68 && 35 && 32 \\
1.3 && 0.9 && 1.44 &&  8.69 && 14.05 && 34 && 28   \\
1.3 && 0.7 && 1.86 &&  8.41 && 11.28 && 33 && 24 \\
1.1 && 0.9 && 1.22 &&  8.41 && 12.27 && 30 && 27 \\
1.1 && 0.7 && 1.57 &&  8.12 &&  9.89 && 29 && 23 \\
1.1 && 0.5 && 2.20 &&  7.81 &&  7.35 && 29 && 18 \\
0.9 && 0.7 && 1.29 &&  7.81 &&  8.41 && 25 && 23 \\
0.9 && 0.5 && 1.80 &&  7.47 &&  6.28 && 24 && 18 \\
0.9 && 0.3 && 3.00 &&  7.10 &&  3.97 && 23 && 13 \\
\hline
\hline
&&&&&& $P_{\rm init}$ = 1.5 d &&&&&& \\
\hline
\hline
1.3 && 1.1 && 1.18 &&  7.38 && 15.16 && 42 && 39 \\
1.3 && 0.9 && 1.44 &&  7.17 && 12.77 && 41 && 35   \\
1.3 && 0.7 && 1.86 &&  6.95 && 10.25 && 41 && 30 \\
1.1 && 0.9 && 1.22 &&  6.95 && 11.15 && 36 && 32 \\
1.1 && 0.7 && 1.57 &&  6.71 &&  8.98 && 36 && 28 \\
1.1 && 0.5 && 2.20 &&  6.45 &&  6.67 && 35 && 22 \\
0.9 && 0.7 && 1.29 &&  6.45 &&  7.65 && 30 && 27 \\
0.9 && 0.5 && 1.80 &&  6.17 &&  5.71 && 29 && 22 \\
0.9 && 0.3 && 3.00 &&  5.86 &&  3.61 && 28 && 16 \\
\hline
\hline
\end{tabular}
\end{table}

The above set of equations was applied to a series of models with different
initial parameters listed in Table~1. The consecutive columns give initial
masses of both components in solar units, initial mass ratio, semi-axis $a$
in solar radii, orbital AM in units of $10^{51}$ gcm$^2$s$^{-1}$ and the
fractional radii of both components relative to the sizes of their Roche
lobes.

As it was demonstrated earlier, the time scale for AML of a binary with the
initial period of 2 d is roughly of the same order as the lifetime of star
``1'' on MS (St\c epie\'n 2006a, 2006b). It was argued by St\c epie\'n
(1995) that 2 d is the expected low limit to orbital period of a close
binary on ZAMS if the binaries are formed in the early fragmentation
process (Boss 1993) accompanied by some AML in the pre-MS phase. The
resulting expected period distribution of young close binaries would be
given by the Duquennoy and Mayor (1991) distribution with the cut-off at 2
d. However, as comparison to observations showed, there is an indication of
an excess of short-period young binaries compared to this prediction (St\c
epie\'n 1995). The excess is even higher for binaries in Hyades (Griffin
1985, St\c epie\'n 1995). No discrepancy was noticed, however, for the
cut-off period in this cluster. The excess of binaries with periods of a
few days can be explained if one notes that majority of them possess a
distant companion. Tokovinin \etal (2006) detected a third component in
96\% of field solar-type binaries with orbital periods shorter than 3 d. A
third component can effectively shorten the close binary period via Kozai
cycles accompanied by the tidal friction (KCTF) if its orbit inclination
exceeds 40$^o$, relative to the orbit of the inner binary (Eggleton and
Kisseleva-Eggleton 2006). After only about 50 Myr the inner orbit can
be circularized in such a case at a period of just a few days, but not less
than 2.5 d. For
shorter periods the mechanism becomes ineffective (see Fig. 2 in Eggleton
and Kisseleva-Eggleton 2006). Similar calculations done by Fabrycky and
Tremaine (2007) showed that KCTF produces predominantly binaries with $P
\approx 2-3$ days. Shorter periods can also be attained by that mechanism
but they need very special conditions and a longer time.  We can therefore
consider the orbital evolution of a typical close binary in a simplified
way as consisting of two separated phases. Binaries with longer orbital
periods (from several up to a few dozen days), which possess the suitably
placed tertiary component, evolve towards shorter periods under the
influence of KCTF on a time scale much shorter than the evolutionary time
scale. AML via magnetized wind, called also magnetic braking (MB), is
negligible for such long periods. When the orbital period reaches a value
of 2-3 days, KCTF mechanism becomes ineffective but MB increases
substantially. The shortening of the period is further continued under the
influence of MB until RLOF and formation of a contact binary. Such a
sequence of both processes in formation of contact binaries is
supported by the ubiquitous presence of third companions to many W~UMa-type
stars observed by Rucinski \etal (2007).

\begin{table}
\caption[]{The final parameters of the computed models}
\begin{tabular}{ccrcccrcccc} \hline\hline

Model && age at RLOF && age/age$_{\rm{TAMS}}$ && $H_{\rm{fin}}$ && $H_{\rm{fin}}/H_{\rm{init}}$ && $P_{\rm{fin}}$ \\
$M_1+M_2(P_{\rm{init}})$ &&  Gyr && && ($\times 10^{51}$) && && days \\
\hline
\hline
1.3+1.1(2.5) && 4.2 && 1.0 && 14.6 && 0.81 && 1.61\\
1.3+0.9(2.5) && 4.2 && 1.0 && 12.3 && 0.81 && 1.61\\
1.3+0.7(2.5) && 4.2 && 1.0 && 9.6 && 0.79 && 1.43\\
1.1+0.9(2.5) && 6.8 && 1.0 && 10.1 && 0.76 && 1.53\\
1.1+0.7(2.5) && 6.8 && 1.0 && 8.1 && 0.76 && 1.43\\
1.1+0.5(2.5) && 6.8 && 1.0 && 5.3 && 0.67 && 0.94\\
0.9+0.7(2.5) && 12.9 && 1.0 && 5.9 && 0.65 && 1.11\\
0.9+0.5(2.5) && 12.9 && 1.0 && 3.8 && 0.56 && 0.67\\
0.9+0.3(2.5) && 8.9 && 0.6 &&  2.1 && 0.49 && 0.35\\
             &&&&&&&&&&\\
1.3+1.1(2.0) && 4.2 && 1.0 && 11.7 && 0.70 && 0.83\\
1.3+0.9(2.0) && 4.2 && 1.0 && 9.9 && 0.68 && 0.83\\
1.3+0.7(2.0) && 4.2 && 1.0 && 7.3 && 0.65 && 0.63\\
1.1+0.9(2.0) && 6.8 && 1.0 && 7.3 && 0.59 && 0.57\\
1.1+0.7(2.0) && 6.6 && 1.0 && 5.7 && 0.58 && 0.52\\
1.1+0.5(2.0) && 5.4 && 0.8 && 4.3 && 0.59 && 0.48\\
0.9+0.7(2.0) && 10.8 && 0.8 && 4.5 && 0.54 && 0.44\\
0.9+0.5(2.0) && 9.1 && 0.6 && 3.5 && 0.56 && 0.42\\
0.9+0.3(2.0) && 6.2 && 0.5 && 2.0 && 0.50 && 0.31\\
             &&&&&&&&&&\\
1.3+1.1(1.5) && 2.7 && 0.7 && 10.8 && 0.71 && 0.62\\
1.3+0.9(1.5) && 2.9 && 0.7 && 8.7 && 0.68 && 0.56\\
1.3+0.7(1.5) && 2.6 && 0.7 && 7.1 && 0.69 && 0.55\\
1.1+0.9(1.5) && 4.1 && 0.6 && 7.1 && 0.64 && 0.47\\
1.1+0.7(1.5) && 4.0 && 0.6 && 5.7 && 0.63 && 0.45\\
1.1+0.5(1.5) && 3.3 && 0.5 && 4.2 && 0.63 && 0.42\\
0.9+0.7(1.5) && 6.3 && 0.5 && 4.5 && 0.59 && 0.38\\
0.9+0.5(1.5) && 5.8 && 0.5 && 3.2 && 0.56 && 0.33\\
0.9+0.3(1.5) && 3.9 && 0.3 && 2.0 && 0.55 && 0.27\\

\hline
\hline
\end{tabular}
\end{table}

Here we are interested in the orbital evolution of close binaries under the
influence of MB so we restrict attention to binaries with short initial
periods. The upper limit is set at 2.5 days. Exploratory model calculations
of binaries with the initial period equal to 3 d show that they reach RLOF
when star ``1'' is already close to the red giant branch that results in
formation of an Algol rather a W~UMa-type star. The lower limit is set at
1.5 days, assuming that a specific combination of higher than average AML
in pre-MS phase and/or special configuration of a tertiary component may
result in such a short initial orbital period. Binaries with the initial
period of 1.5 d will reach RLOF in a shorter time than binaries with the
longer initial periods and form contact binaries with a relatively young
age. We know at least one W~UMa-type star with the age substantially lower
than the limiting age of 4-5 Gyr found by Kaluzny and Rucinski (1993), and
Rucinski (1998, 2000). It is TX Cnc -- a member of an intermediate-age open
cluster Praesepe with the age of only 600 Myr (Zhang, Deng and Lu 2009). A
few more W~UMa-type stars show kinematically young age (Bilir et
al. 2005). Dynamical properties of an individual star do not describe
directly its true age but some of the kinematically young stars may in fact
be also evolutionary young. We may conclude that young contact stars are
observed although they are quite rare.  The initial masses of star ``1''
are taken from the range between 1.3 $M_{\odot}$ and 0.9 $M_{\odot}$ to
cover the range from the adopted upper limit for the existence of the
subphotospheric convection zone, down to the mass for which lifetime on MS
is shorter than the Hubble age. The initial masses of star ``2'' are fitted
to avoid extreme mass ratios \ie twin binaries, or very low mass ratio
binaries in which the rapid mass transfer following RLOF is
nonconservative.  The considered initial mass ratios are between 1.18 and
3.0. The initial orbital AM range extends from 3.6 to 17 in units of
$10^{51}$ gcm$^2$s$^{-1}$. As we see from the two last columns of Table~1,
the modeled binaries are initially well detached, with the radii of star
``1'' between 20\% and 42\%, and the radii of star ``2'' between 11\% and
39\% of the size of their respective Roche lobes. All binaries are assumed
to have zero eccentricity and fully synchronized rotation of both
components with the orbital period.

\begin{figure}[htb]
\centerline{\includegraphics[width=12.5cm]{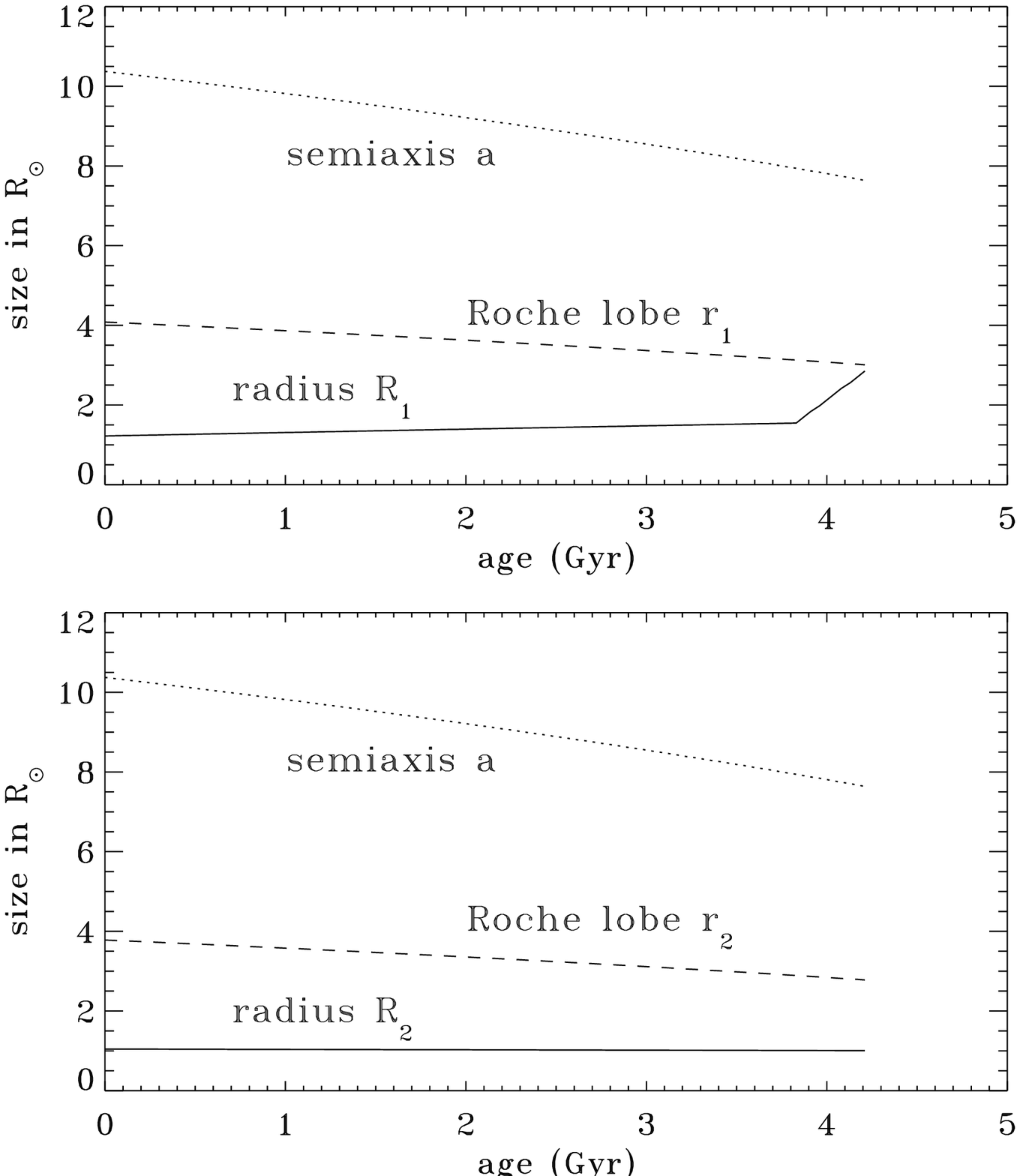}}
\FigCap{Time variations of geometrical parameters of the binary with
initial masses 1.3+1.1 $M_{\odot}$ and initial period of 2.5 d. Star ``1'' 
reaches TAMS when the binary is still detached.}
\end{figure}

\begin{figure}[htb]
\centerline{\includegraphics[width=12.5cm]{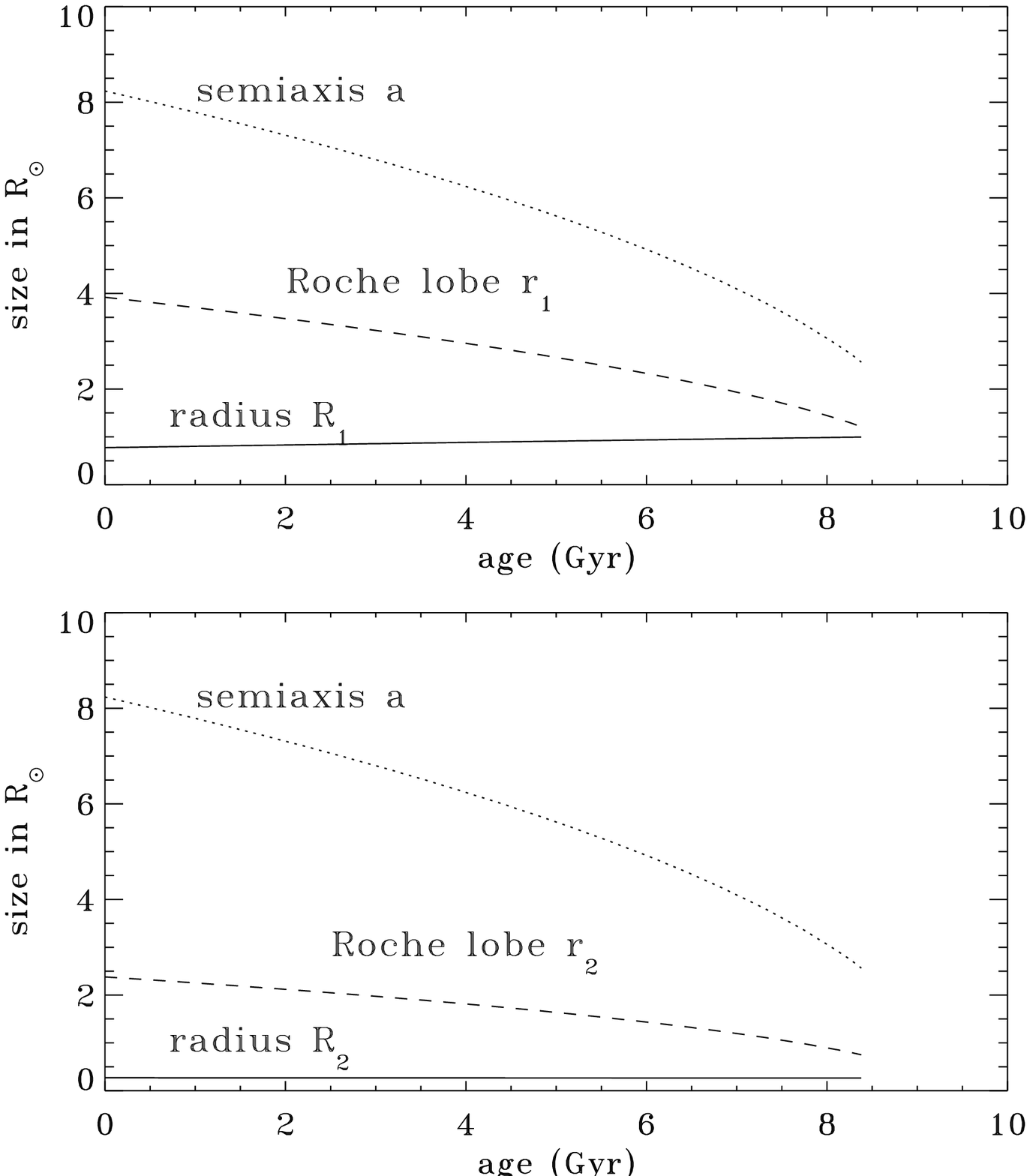}}
\FigCap{Time variations of geometrical parameters of the binary with
initial masses 0.9+0.3 $M_{\odot}$ and initial period of 2.5 d. Here star
``1''  reaches the critical Roche lobe when it is still on MS.}
\end{figure}

\section{Results of the modeling}
\subsection{Time evolution of the orbital period}

Eqs. (7-8) are the basic equations governing the evolution of geometrical
parameters of a close binary in the detached phase. They are integrated in
time, starting from initial conditions. The amount of mass and orbital AM
lost at each time step is calculated and subtracted from the corresponding
quantities. Radii of components are calculated from a grid of
evolutionary models of single stars with the solar composition 
by simple interpolation
formulas. Models obtained by Schaller \etal(1992), Girardi \etal (2000)
and models of very low mass stars obtained by Dr. R. Sienkiewicz for the
present author were used.\footnote{A description of the latter models is
given in St\c epie\'n (2006a) and the details of the modeling program are
given in Paczy\'nski \etal (2007).} New values of masses, radii and
$H_{\rm{orb}}$ are used to calculate all other binary parameters at each
time step. The calculations stop when star ``1'' fills its inner critical
surface. Depending on initial conditions, it happens when the star is still
evolving across MS, when it is at terminal age MS (TAMS) or beyond it.
Following RLOF mass exchange between the components takes place in case A,
AB or B, respectively. The subsequent evolution of the binary depends
sensitively on the evolutionary advancement of star ``1'' at the time of
RLOF. This evolution will be the subject of the future investigation. At
present we concentrate on the early stages of evolution, when the binary
is still detached.

Table~2 summarizes the results of the model calculations. All columns are
self explanatory. 
Two characteristic evolutionary models are presented in Figs.~1 and
2. Fig.~1 shows the time variations of geometrical parameters of the most
massive binary with the initial masses 1.3+1.1 $M_{\odot}$ and the longest
initial period of 2.5 d. The binary has also the highest initial AM (see
Table~1). After about 4 Gyr star ``1'' approaches TAMS when its radius is
still significantly smaller than the size of the Roche lobe. The period has
shortened to the value of 1.6 d during this time. When hydrogen is
exhausted in the center, the star expands quickly and fills its Roche
lobe. Fig.~2 shows the same data for the least massive binary from Table~2
with initial masses 0.9+0.3 $M_{\odot}$ and the initial period of 2.5
d. Here we see that the Roche lobe descends onto the surface of star ``1''
when it is still on MS. The binary has a period of only 0.35 d at that
time.

\begin{figure}[htb]
\centerline{\includegraphics[width=12.5cm]{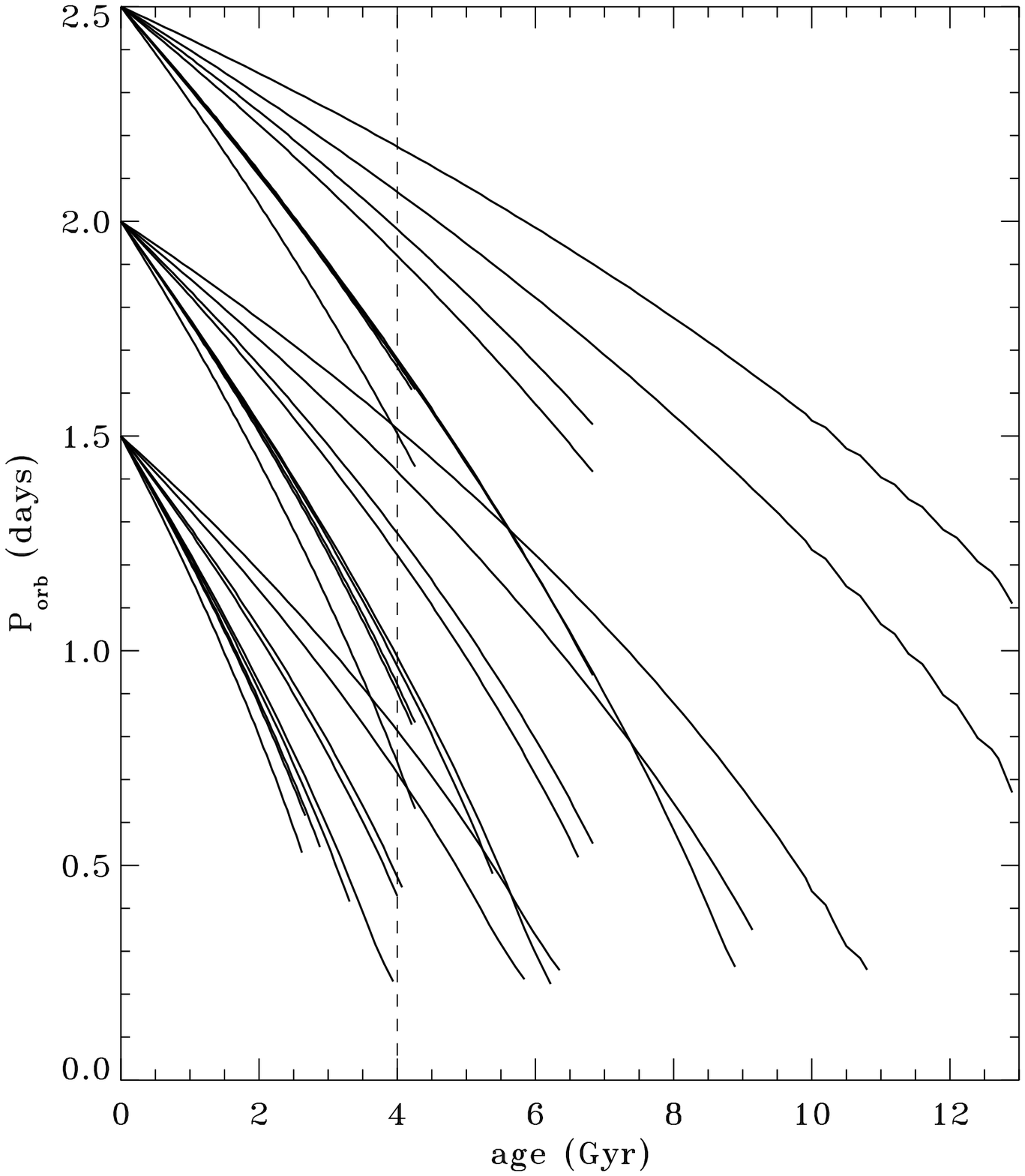}}
\FigCap{Time variations of orbital periods of all the considered
binaries. The initial and end coordinates of the plotted curves are given
in Table~1 and Table~2, respectively.}
\end{figure}

The period variations of all the considered models are presented in
Fig.~3. All the plotted curves show a bend increasing with age. This is a
consequence of the period dependence of the AML rate (see Eq.~7 above) -
the shorter the period, the higher AML rate, hence the faster period
shortening. The vertical broken line marks the limiting age of 4 Gyr beyond
which stellar clusters contain high numbers of W UMa-type
variables. Clusters younger than that age contain no, or very few such
stars. % (Kaluzny \& Rucinski 1993, Rucinski 1998, 2000). 
Almost all the plotted curves reach that line and many go beyond. This
indicates that very few binaries will lose enough AM for RLOF to occur in
time shorter than 4 Gyr. As the data in Table~2 show, this happens only for
binaries with the initial orbital period equal to 1.5 d and mass of star
``1'' equal to 1.3 $M_{\odot}$. The mass of star ``2'' does not matter in
this case. For the mass of star ``1'' equal to 1.1 $M_{\odot}$ RLOF occurs
before 4 Gyr only when the initial mass of star ``2'' is equal to 0.5
$M_{\odot}$ (see Table~2). However, all binaries with $P_{\rm init}$ =
1.5 d reach RLOF when star ``1'' is only about halfway to TAMS. Star ``2''
is even less evolutionary advanced. After the mass exchange, 
many such binaries may not fulfill the
geometrical condition of the Roche model required to form a contact
binary. Instead, a near contact binary will be formed in such a case, 
with star ``1'' filling its critical Roche surface and slowly
transferring matter to star ``2'' which is still within its Roche lobe. The
results of the calculations also show that the initial period of TX Cnc
must have been even shorter -- of the order of 1 d. Binaries with very
short periods can be formed in stellar clusters by the interaction with
other cluster members (\eg Hut 1983) although paucity of such binaries in
young and intermediate age clusters indicates that such events are not
frequent.

\begin{figure}[htb]
\centerline{\includegraphics[width=12.5cm]{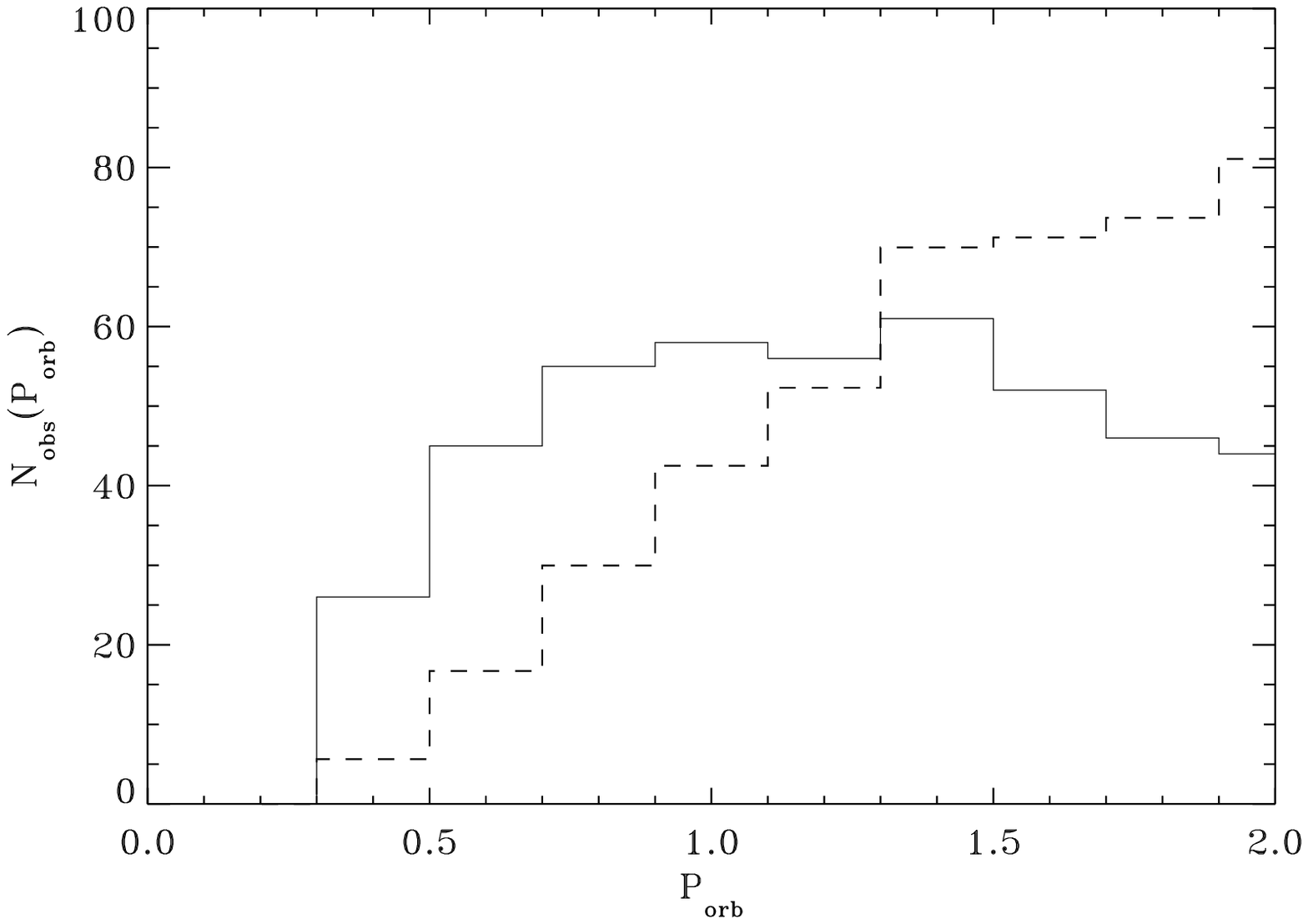}}
\FigCap{The uncorrected observed period distribution, based on ASAS data
for 421 
detached binaries (solid line), compared to the corrected distribution,
normalized to the same total number of binaries as observed 
(dotted line). See text for details.}
\end{figure}

The initial period of 2.5 d is long enough for nearly all stars ``1'' in
the considered binaries to reach TAMS before MB leads to RLOF. In other
words, both components are still well within their respective Roche lobes
when star ``1'' completes its MS evolution and accelerates expansion on its
way to the red giant branch. There is only one exception from this:
 the least massive model 0.9+0.3(2.5). Here the Roche lobe
descends onto the surface of star ``1'' after about 9 Gyr (see Table~2)
when it is still on MS. 
In models with $P_{\rm init}$ = 3 d the AML rate is even lower
than in models with $P_{\rm init}$ = 2.5 d. As a result, in all cases, star
``1'' reaches TAMS when the period is barely shortened to values close to 2
d. Mass exchange in such binaries takes place in case B, resulting in
Algols with periods of a few days which are beyond the scope of the present
paper.

\subsection{Period distribution}

It was argued in Section 2 that the KCTF mechanism can effectively produce
close, detached binaries with periods of a few days, in a time much shorter
than the evolutionary time scale (Eggleton and Kiseleva-Eggleton 2006,
Fabrycky and Tremaine 2007). As a result, an excess of close binaries with
such periods should occur among young stars, compared to expectations based
on the standard period distribution (Duquennoy and Mayor 1991). The
observations of Hyades show indeed an existence of such an excess (Griffin
1985, St\c epie\'n 1995). The crucial role of the KCTF mechanism in
producing this excess is strongly suggested by the recent analysis of field
binaries by Tokovinin \etal (2006) which showed that nearly all binaries
with periods shorter than 3 d possess a distant companion. The effectiveness
of the KCTF mechanism drops, however, rapidly for periods shorter than
2-2.5 d. The resulting period distribution produced solely by KCTF should
show a cut-off at this value. A similar cut-off period of 2 d is expected
for ``pure'' close binaries formed in the early fragmentation process (for
a discussion see St\c epie\'n 1995). The cumulative initial period
distribution of young cool binaries is therefore expected to have a cut-off
near 2 d. Due to the magnetized winds, orbits of such binaries evolve further
towards shorter periods but this evolution is now determined by MB. 
The period distribution predicted by model calculations can be
determined and compared with the observed distribution.

To obtain the observed period distribution, the data from Pilecki (2010)
were used. He selected a sample of short-period eclipsing binaries from the
ASAS survey (for information see http://www.astrouw.edu.pl/asas) brighter
than $11^{\rm m}.5$ and frequently observed in two ($V$ and $I$)
colors. Then he solved their light curves with the Wilson-Devinney code and
obtained the basic binary parameters. We selected binaries marked as
detached by Pilecki, with periods shorter than 2 d, and with temperatures
from the interval 5200-6600 K, corresponding approximately to the mass of
the more massive component between 0.9 and 1.3 $M_{\odot}$. The total
number of such binaries is 421. The plot of the actual numbers in bins with
the width of 0.2 d is given in Fig.~4 as a solid line. The distribution
based on eclipsing binaries needs a correction for the bias connected with
the orbit inclination. A simple estimate of the correction factor is
derived by Maceroni \& Rucinski (1999). According to them, the number of
stars observed at each period should be multiplied by $P^{4/3}$. Thus
corrected distribution, normalized to the total observed number of
binaries, is plotted in Fig.~4 as a broken line. The latter distribution
will be compared to the predicted distribution resulting from the model
computations.

The predicted period distribution is calculated under the following
assumptions:

\begin{enumerate}

\item Stationary state prevails \ie the volume density of binaries at each
value of the orbital period is constant.

\item Initial period and component masses are not correlated, \ie 
      $n(P,M_1,q)dPdMdq$ = $n_1(P)n_2(M_1)n_3(q)dPdMdq$, where $n$ is the number density of
   binaries with different periods and component masses whereas $n_1, n_2,
   n_3$ are the number densities depending on period, mass and mass ratio,
   respectively.

\item Mass variations of both components are neglected during the
   considered time so $n_2$ and $n_3$ are time-independent. The
   only time variable is $n_1$.

\item The initial period distribution $n_1(P_{\rm init})$ = const. down to
   the cut-off period. Two values of the cut-off period are considered: 
1.5 d or 2 d.

\item The mass distribution is given by the initial mass function of
   Salpeter (1955):  ${\rm d}n_2(M_1) \propto M_1^{-2.35}{\rm d}M_1$

\item The mass ratio distribution $n_3(q)$ = const. 

\end{enumerate}

\begin{figure}[htb]
\centerline{\includegraphics[width=12.5cm]{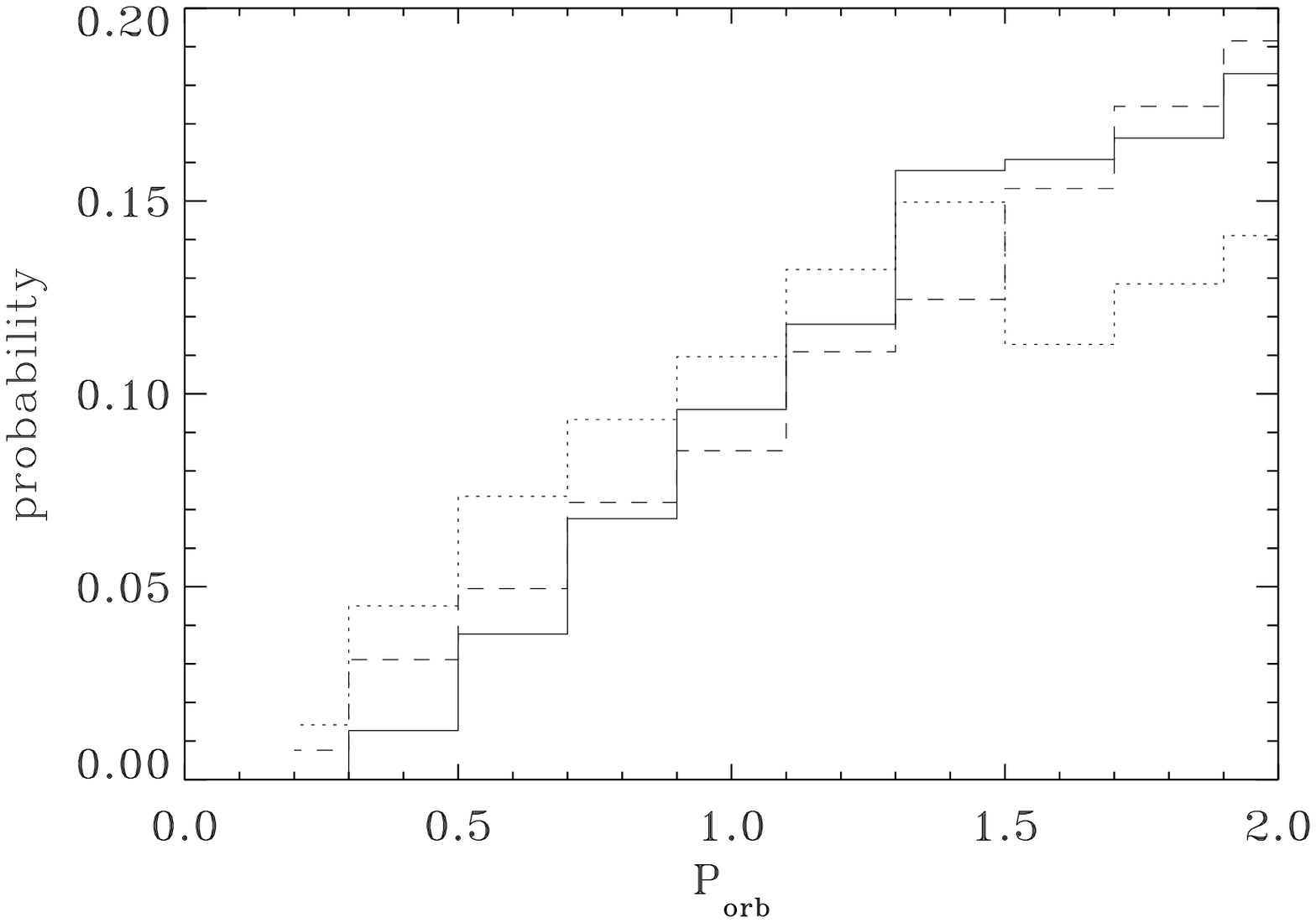}}
\FigCap{The observed period distribution of the detached binaries, 
based on the ASAS data and corrected
for the detection bias as described in the text, is shown by the  
solid line. It is compared to the predicted distribution with the cut-off
period equal to 1.5 d (dotted line) and to 2 d (broken line).}
\end{figure}

Let us consider the evolution of period along a given curve $P_i(t)$, see
Fig.~3. In the stationary state, the number density $n_i$ of binaries
evolving along that curve and having periods
between $P$ and $P+\Delta P$ is proportional to the relative time spent by a
binary within this period interval, \ie

\begin{equation}
n_i(P, P+\Delta P)= \frac{1}{T_i} \int_{t}^{t+\Delta t}\frac{{\rm d}t}{{\rm
d}P_i/{\rm d}t}\,,
\end{equation}

where $T_i$  is the total evolutionary time (listed in column 2 of Table~2).

To obtain the final expected period distribution, $n_1$, a summation of all
$n_i$ for the respective period interval should be done, with weights
resulting from assumptions (4)-(6) above

\begin{equation}
n_1(P, P+\Delta P) = \sum_{i}\frac{T_iw_in_i(P, P+\Delta P)}{T}\,,
\end{equation}

where $w_i$ are the weights
and $T = \sum T_i$. Summation was done over all models listed in Tables~1
and 2. The resulting distributions with the cut-off period equal to 1.5 d
(dotted line) and to 2 d (broken line) are compared in Fig.~5 to the
observed distribution, corrected for the selection bias. Both predicted
distributions were binned in the same way as the observed one. All three
distributions are normalized to the unity area. It is immediately visible
that the predicted distribution with the cut-off period of 2 d fits the
observational data much better than the one with the cut-off period of 1.5
d. The results of the $\chi^2$ - test support this conclusion. The value of
the reduced $\chi^2$ is equal to 1.04 in the first case and to 2.91 in the
latter case. One more predicted distribution was also calculated, with a
weight of 1/3 given to all binaries with initial period of 1.5 d, compared
to unity given to binaries with 2 and 2.5 d (in the two previous cases this
weight was equal to 1 and 0, respectively). That distribution is not
plotted in Fig.~5 to avoid line crowding but the value of the reduced
$\chi^2$ in this case is equal to 1.44 which indicates that the
distribution is acceptably close to the observed one although the agreement
is worse than in case of the distribution with the cut-off period of 2
d. This result suggests that the pool of short period binaries evolving
further under the influence of MB may also contain a small share of
binaries with initial periods of 1.5 d.

\subsection{Uncertainties of the model} 

The main uncertainties of the presented model are connected with
inaccuracies of the numerical coefficients in Eqs.~(7-8) describing the AML
rate and the mass loss rate. These inaccuracies do not alter significantly
the main qualitative conclusions, however, some of the quantitative results
may change with a change of the coefficients. For example, if we increase
the AML rate by the uncertainty limit \ie 50 \%, the time till RLOF will
shorten correspondingly. To obtain the final values similar to those listed
in Table~2 the initial periods of all considered binaries would have to be
increased by a few tenths of a day. A decrease of the AML rate would result
in lengthening of the time till RLOF and an increased evolutionary
advancement of the stars filling their Roche lobes. To obtain the values
close to those listed in Table~2 the initial periods of all considered
binaries would have to be {\it decreased} by a few tenths of a
day. Allowing for these uncertainties it is concluded that the detached
cool binaries with initial periods between 1.5 and 3 d are the progenitors
of short period semi-detached and contact binaries.  An increase/decrease
of the mass loss rate would increase/decrease the MS life times of the more
massive components and decrease/increase the final component masses by a
few percent when RLOF occurs, compared to the values given in the paper.
This will modify to some degree the further evolutionary stages of the
binaries.

Another source of uncertainties results from the use of single star
models. A comparison of the existing grids of models reveals differences
among them at
the level of several percent in stellar radii, MS life times etc. They
result from different input physics and computational methods. In addition,
the observed radii of cool MS stars are systematically larger by about 10
\% than obtained from models (Torres \etal 2010) which is probably
connected with the presence of magnetic fields in these stars. A similar
uncertainty of several percent in stellar parameters is introduced by
neglecting the influence of the component interaction on their
evolution. It is rather difficult to precisely calculate the influence of
model inaccuracies on the presented results but we expect that the main conclusions of the present study 
are not altered.

\section{Main Conclusions}

A set of 27 evolutionary models of cool close binaries with the initial
periods of 1.5, 2.0 and 2.5 d was computed under the assumption that mass
and AM loss due to the magnetized winds of both components determine the
orbital evolution until the more massive component fills its critical Roche
lobe. Following RLOF, mass transfer between the components occurs which
results in formation of a semi-detached or contact binary with the mass ratio
reversed. Nine models were computed for each period with different
component masses (see Table~1). The present paper deals with the first
phase of binary evolution - from the beginning till RLOF. The consecutive
phases which include mass transfer following the RLOF and
the further evolution towards contact or a classical Algol configuration
will be the subject of the subsequent study.

An important test for the correctness of the adopted assumptions and model
calculations results from the comparison of the predicted period
distribution to the observed distribution of detached binaries with the
same masses. The comparison was restricted to periods shorter than 2 d
where, according to the model assumptions, the period evolution is
determined by MB. An excellent agreement is obtained for the flat
initial period distribution with a cut-off at 2 d. A flat distribution with
a cut-off at 1.5 d differs significantly from the observations, however, a
small admixture of such stars is not excluded. This result indicates that
MB is a dominating mechanism populating the orbital period region below 2
d. Other mechanisms, like KCTF or N-body interactions play a minor role in
producing binaries with such short periods.

\begin{figure}[htb]
\centerline{\includegraphics[width=12.5cm]{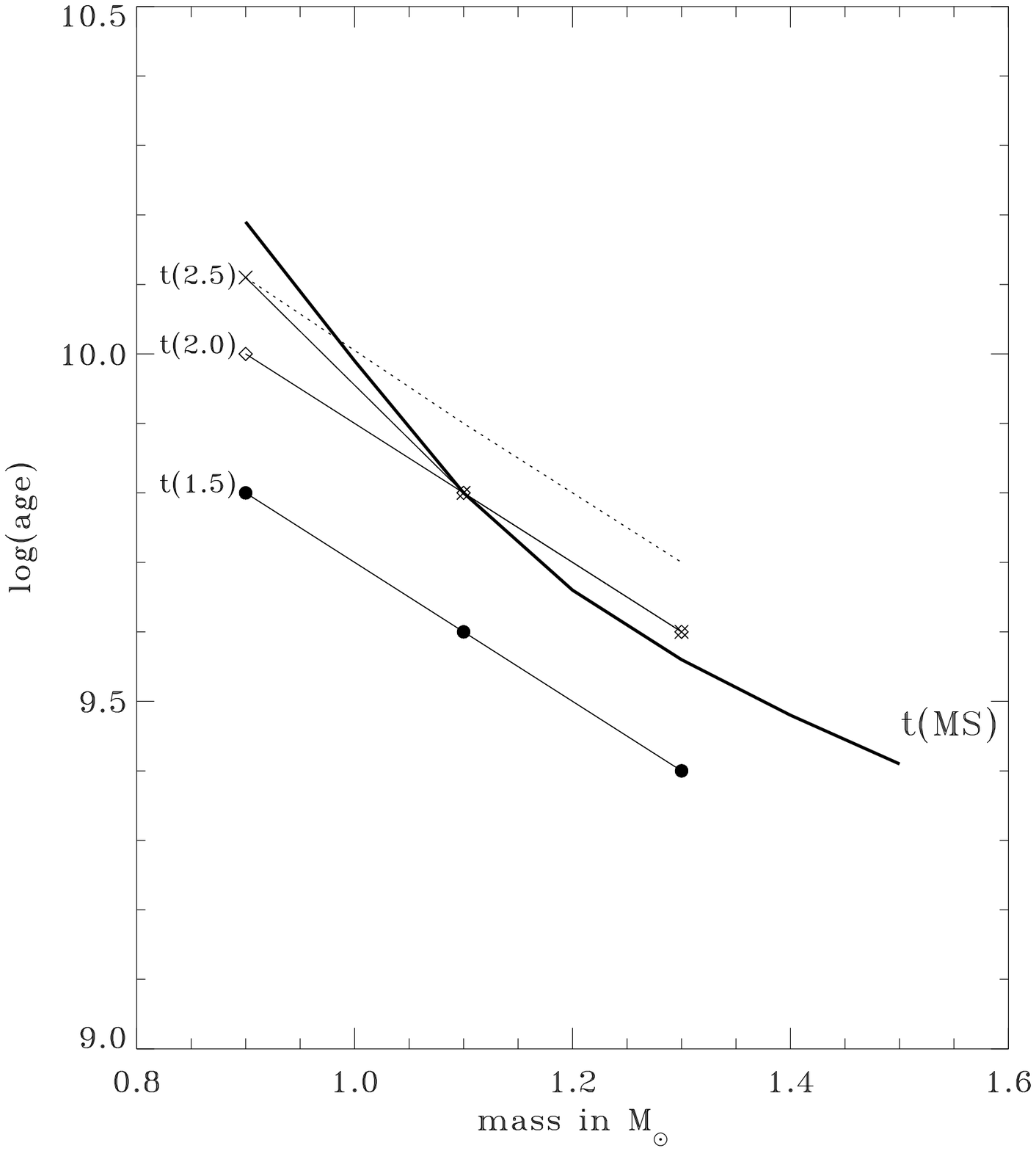}}
\FigCap{Comparison of the MS life time of the massive component
(heavy solid line) with time needed to reach RLOF for binaries with 
different initial orbital
periods (light solid lines). A broken line gives time needed for the
occurrence of RLOF in a
fictional situation when only MB operates and the star does not expand
beyond its TAMS size.}
\end{figure}

The results show that it is the length of the initial orbital period
that primarily controls the evolution of a cool close binary in phases
following the rapid mass exchange. As it was mentioned earlier, RLOF occurs
due to the operation of two, equally important processes: the evolutionary
expansion of the massive component of the binary and the contraction of its
Roche critical lobe resulting from MB. Fo the discussed binaries time scales of both processes are
roughly of the same order. However, while the MS life time of a star does not
depend on the orbital parameters, the MB time scale depends sensitively on
the initial length of the orbital period (and also, to some degree, on the
initial mass ratio). This is illustrated in Fig.~6 where the MS life time
of the massive component, plotted as a heavy solid line, is compared to the
time needed for RLOF to occur in binaries with different initial orbital
period. The latter times are shown as light solid lines labeled with a value of
the initial period. MS life times refer to the Girardi
\etal (2000) models with solar composition. The times to reach RLOF were taken
from Table~2 for binaries with moderate mass ratios. Note that they cannot
significantly exceed MS life time because the star expands beyond TAMS
quickly until it fills the Roche lobe. A broken line describes the time needed
for the occurrence of RLOF in binaries with the initial period of 2.5 d,
assuming the fictional situation that the massive component does not
expand beyond its TAMS size and the Roche lobe descends onto the stellar
surface solely due to MB.

It is seen from Fig.~6 that the MB time scale for binaries with the initial
period of 2.5 d or more is longer than the MS life time of the massive
component that reaches TAMS being well inside its Roche lobe whereas the
binary still possesses a substantial fraction of its initial AM. As a
result, the star expands and fills the Roche lobe on its way to the red
giant branch. Following the rapid mass exchange (in the so called case B) a
short period Algol is formed in which the (slow) mass transfer prevails
over MB and the orbital period increases. Only for binaries with the
massive component of 0.9 $M_{\odot}$ the time to reach RLOF is shorter than
the MS life time but this is a result of the solar composition which was
used in all calculations to preserve consistency. The 0.9 $M_{\odot}$ model
with low metallicity, characteristic of globular clusters would be more
appropriate here. Its MS lifetime is around 10 Gyr as compared to about 15
Gyr for a solar composition star. The lifetime of 10 Gyr is shorter than
the MB time scale of a binary with the initial period of 2.5 d and is close
to the time for RLOF occurrence in a binary with the initial period of 2 d
(see Fig.~6).

For binaries with the initial period of 2 d the MS life times are close to
the time needed for RLOF to occur so the massive component
fills the Roche lobe just when it is at, or close to TAMS and the case AB
mass transfer follows. The orbital period shortens to only a fraction
of a day at that point and the rapid mass exchange results in a formation
of contact, or near contact binary which soon reaches full contact. During
the further evolution the (slow) mass transfer approximately balances AML
and the orbital period does no vary much when the mass ratio decreases to
its extreme value of about 0.1.

For binaries with the initial period of 1.5 d or less the time to reach
RLOF is shorter than the MS life time of the massive component. The star
fills the Roche lobe during its MS evolution and the case A mass transfer
follows. The orbital period is very short at RLOF - typically less than 0.5
d.  After the mass exchange a contact or near contact binary is
formed. Because MB prevails over the (slow) mass transfer, the orbital
period shortens and the coalescence of both components occurs for a
moderate rather than extreme mass ratio.

The present calculations cannot be directly applied to globular clusters
as the stellar evolutionary models with low metallicity should be used in
modeling evolution of cluster binaries. Nonetheless it is possible to sketch a
qualitative picture regarding the presence of contact binaries in these
clusters.

Members of a globular cluster reach TAMS when they are at the turn-off
point. If such a star is also a member of a close binary with a lower mass
component and a proper initial orbital period (around 2 d, although more
accurate value will follow from a full scale evolutionary calculations),
RLOF occurs, followed by the approximate mass ratio inversion. A contact
binary is formed in which the presently more massive component has a mass
close to the pre-RLOF massive component, \ie close to the turn-off mass.
The new-born contact binaries are therefore expected to appear in the
turn-off region. During the subsequent evolution towards the extreme mass
ratio the contact binaries move to the region of blue stragglers until the
ultimate coalescence occurs. Contact systems can also be formed below the
turn-off region from binaries with initial periods significantly shorter
than those of contact binaries from the turn-off region. An apparent
shortage of such systems in the observed clusters indicates that mechanisms
producing cool binaries with very short initial periods are not very
efficient in globular clusters, similarly as it is in open clusters or
among field binaries.

The above picture is very simplified and approximate. It will vary in
actual situations from one binary to another, depending on initial values of
individual component masses, their chemical composition, orbital
parameters and details of the evolutionary modeling.

The final values from Table~2 will be used as input data for calculating
the subsequent evolution of the binaries. The results will be compared with
the observations of W UMa-type contact binaries and related stars.

\Acknow{I thank Slavek Rucinski for a careful reading of the manuscript and
several remarks which substantially improved the paper.}

\end{document}